\pdfoutput=1
\documentclass[sigconf]{acmart}

\usepackage{color}
\usepackage{bbm}
\usepackage{multirow}
\usepackage[inline]{enumitem}
\usepackage{subfigure} 
\usepackage{sistyle}
\usepackage[ruled]{algorithm2e}
\usepackage{booktabs}
\usepackage{caption}
\SIthousandsep{,}
\usepackage{makecell}
\usepackage[skip=0pt]{caption}
\usepackage{amsmath}
\usepackage{hyperref}
\usepackage{array} 
\usepackage{graphicx}
\usepackage{xcolor}
\usepackage{acronym}

\usepackage{amssymb}
\usepackage[export]{adjustbox}
\usepackage{colortbl}
\definecolor{lightgray}{gray}{0.9}
\newcommand{\heading}[1]{\vspace*{0.5mm}\noindent\textbf{#1.}}

\AtBeginDocument{%
  \providecommand\BibTeX{{%
    \normalfont B\kern-0.5em{\scshape i\kern-0.25em b}\kern-0.8em\TeX}}}

\makeatletter
\g@addto@macro\normalsize{%
  \abovedisplayskip 2pt plus1pt 
  \belowdisplayskip 2pt plus1pt
  \abovedisplayshortskip  2pt plus1pt%
  \belowdisplayshortskip  1pt plus1pt
}
\makeatother

\newcommand{\Vanilla}{Vanilla LLM\xspace}
\newcommand{\RAG}{T-RAG\xspace}
\newcommand{\PRAG}{P-RAG\xspace}
\newcommand{\DyPRAG}{DyP-RAG\xspace}
\newcommand{\PRAGCombine}{PT-RAG\xspace}
\newcommand{\DyPRAGCombine}{DyPT-RAG\xspace}

\newcommand{\Dataset}{News25-QA\xspace}

\acrodef{IR}{information retrieval}
\acrodef{LLM}{large language model}

\AtBeginDocument{%
  \providecommand\BibTeX{{%
    Bib\TeX}}}

\setcopyright{acmlicensed}
\copyrightyear{2018}
\acmYear{2018}
\acmDOI{XXXXXXX.XXXXXXX}

\acmConference[Conference acronym 'XX]{Make sure to enter the correct
  conference title from your rights confirmation emai}{June 03--05,
  2018}{Woodstock, NY}

\acmISBN{978-1-4503-XXXX-X/18/06}


\author{Minghao Tang}
\orcid{0009-0002-1911-5142}
\affiliation{
    \institution{State Key Laboratory of AI Safety,}
    \institution{ICT, Chinese Academy of Sciences}
	\institution{University of Chinese Academy of Sciences}
	\city{Beijing}
	\country{China}
}
\email{tangminghao25@mails.ucas.ac.cn}

\author{Shiyu Ni}
\orcid{0009-0001-7965-7771}
\affiliation{
    \institution{State Key Laboratory of AI Safety,}
    \institution{ICT, Chinese Academy of Sciences}
	\institution{University of Chinese Academy of Sciences}
	\city{Beijing}
	\country{China}
}
\email{nishiyu23z@ict.ac.cn}

\author{Jingtong Wu, Zengxin Han}
\affiliation{
    \city{}
	\country{}
}
\email{wangzhenlingwu@163.com}
\email{zengxin.hanzx@gmail.com}

\author{Keping Bi}
\orcid{0000-0001-5123-4999}
\affiliation{
    \institution{State Key Laboratory of AI Safety,}
    \institution{ICT, Chinese Academy of Sciences}
	\institution{University of Chinese Academy of Sciences}
	\city{Beijing}
	\country{China}
}
\email{bikeping@ict.ac.cn}

\renewcommand{\shortauthors}{Minghao Tang et al.}
\begin{document}

\title{Understanding Parametric Knowledge Injection in Retrieval-Augmented Generation}

\begin{abstract}
Context-grounded generation underpins many LLM applications, including long-document question answering (QA), conversational personalization, and retrieval-augmented generation (RAG). However, classic token-based context concatenation is costly for long inputs and can be ``lost in the middle'' at extreme context lengths. Recent work explores \textit{context parameterization}, which encodes context into lightweight trainable parameters (e.g., LoRA adapters) injected into a frozen LLM. Extending this idea to retrieved evidence yields \textbf{parametric RAG} (P-RAG), which incorporates knowledge via parameter updates rather than token-level attention.

In this paper, we present a systematic study of this emerging RAG paradigm—parametric knowledge injection. First, we reassess P-RAG under answer-presence accuracy and show that it does not consistently outperform standard token-based RAG (T-RAG), while combining both (PT-RAG) achieves the best overall performance. Second, we introduce a QA benchmark with up-to-date knowledge beyond the LLM's internal memory to enable controlled analysis. Our representational and mechanistic results indicate that parametric representations capture document-level semantics and primarily influence deeper feed-forward computations, providing high-level guidance but limited evidence consolidation. Finally, we evaluate parametric injection under key RAG challenges, demonstrating improved faithfulness under knowledge conflicts, stronger robustness to retrieval noise, and solid generalization to tasks beyond QA. Our findings clarify the strengths and limitations of parametric RAG and provide practical guidance for future retrieval-augmented LLM systems.

\end{abstract}


\maketitle
\acresetall

\section{Introduction}
Context-grounded generation has become a standard paradigm for many large language model (LLM) tasks, including long-document question answering (QA) over books or reports~\cite{kovcisky2018narrativeqa, dasigi2021dataset}, leveraging long-term and short-term conversational history for personalization and consistency~\cite{packer2023memgpt, zheng2025lifelong}, in-context learning from a few exemplars~\cite{brown2020language, dong2024survey}, and retrieval-augmented generation (RAG)~\cite{lewis2020retrieval, izacard2023atlas}. 
In most existing systems, an LLM incorporates context by directly ingesting the context tokens—analogous to instructions—and attending over them during generation. However, this attention-based mechanism incurs $O(l^2)$ computational complexity with respect to the context length $l$, which makes long-context processing expensive and often forces practical context size constraints. 
While recent frontier models such as Claude and Gemini have extended their context windows to as many as one million tokens~\cite{gemini-3-pro,claude-sonnet-4.5}, LLMs still struggle to reliably utilize very long inputs and can become ``lost in the middle''~\cite{liu2024lost}. Therefore, effectively and efficiently leveraging contextual information remains a central research challenge for context-grounded LLM generation.

To address these challenges, recent research has explored context \textbf{parameterization}~\cite{wang2024greater, cao2025infiniteicl, chengenerative, wangself, liao2025e2llm}, which encodes context into a small set of trainable parameters (e.g., soft prompts or LoRA adapters). These parameters are plugged into a frozen LLM and optimized using context-related objectives, typically supervised next-token prediction or QA-style training. At inference time, the model can access the contextual information through these internal parameters, rather than explicitly attending to external context tokens. This enables long contexts to be incorporated beyond the model’s context window limit, while substantially reducing inference cost since the input sequence only contains the task instruction.

In retrieval-augmented generation (RAG), context parameterization has recently been extended to incorporate retrieved evidence, giving rise to \textbf{P}arametric \textbf{RAG} (\textbf{P-RAG})~\cite{su2025parametric}. 
Beyond improving efficiency and mitigating long-context bottlenecks, P-RAG aims to integrate external knowledge in a way that influences generation similarly to the model's internal (parametric) knowledge. 
Concretely, P-RAG converts retrieved evidence into lightweight parameter updates (e.g., via LoRA adapters), injecting knowledge through the model’s feed-forward computations rather than via explicit attention over context tokens, as in classic \textbf{T}oken-based \textbf{RAG} (\textbf{T-RAG}). 
Since prior work suggests that factual associations are largely captured in the feed-forward layers of LLMs~\cite{geva2021transformer}, P-RAG may provide a more direct pathway for knowledge incorporation, potentially benefiting settings that require multi-step reasoning and consolidation across evidence. 
Empirically, \citet{su2025parametric} show that P-RAG outperforms T-RAG on several multi-hop QA benchmarks, and that combining parametric and token-based context yields further gains. Motivated by these results, the shift from token-based conditioning to parametric knowledge injection in RAG has attracted increasing attention~\cite{tan2025dynamic, chen2025privacy, su2025routing, fan2025one, su2025dynamic, fleshman2025lora}.



Despite its potential, P-RAG also introduces several limitations compared to token-based RAG (T-RAG). First, parametric representations inevitably compress the original text and may lose fine-grained details. 
Second, they are less interpretable than explicit token-level evidence. 
Third, the learned adapters are typically model-specific, meaning they can only be applied to the same LLM checkpoint (or closely related variants) on which they were trained. 
Fourth, parametric representations can be sensitive to the training procedure and hyperparameter choices, which may affect stability and reproducibility. 
In addition, training and storing separate LoRA adapters for each retrieved passage can be costly. More comparisons between P-RAG and T-RAG are summarized in Table~\ref{tab:compare_RAG}.


Furthermore, existing P-RAG studies often rely on F1 \cite{su2025parametric,tan2025dynamic, chen2025privacy}, which combines precision and recall, rather than the commonly used HasAnswer accuracy that only checks whether the generated response contains the ground-truth answer or a consistent equivalent.
This can be misleading because precision is highly sensitive to response length: among correct responses, shorter outputs tend to achieve higher precision, which may systematically favor P-RAG due to its typically more concise generations. Therefore, it is necessary to reassess P-RAG using the HasAnswer metric, for both overall performance evaluation and further analysis. Figure~\ref{fig:case_study} illustrates how these metrics can lead to different conclusions.

To better understand parametric knowledge injection in RAG and provide practical guidance for future work adopting the P-RAG paradigm, we conduct a systematic study of parametric RAG (focusing on P-RAG~\cite{su2025parametric} and its extension DyP-RAG~\cite{tan2025dynamic}) from multiple perspectives. 
Specifically, we aim to answer the following research questions (RQs):




(\textbf{RQ1}) \emph{Evaluation validity.} Do P-RAG methods still outperform T-RAG when evaluated with answer-presence accuracy (HasAnswer), rather than length-sensitive F1?

(\textbf{RQ2}) \emph{Representation and mechanism.} Do parametric representations encode passage semantics effectively, and how does parametric injection affect the LLM’s computation? Specifically, does it preserve passage semantics, alter internal activations in a meaningful way, and improve multi-hop reasoning and evidence consolidation?

(\textbf{RQ3}) \emph{Context faithfulness, robustness, and generalization.} How does P-RAG behave under core RAG challenges, including (i) faithfulness under conflicts between externally retrieved and internal parametric knowledge, (ii) robustness to irrelevant or noisy evidence, and (iii) generalization to tasks beyond QA?

For \textbf{RQ1}, we find that P-RAG does not consistently outperform T-RAG under the HasAnswer metric. However, consistent with prior findings, combining P-RAG with T-RAG (i.e., PT-RAG) achieves the best performance across settings. This suggests that parametric injection alone may not sufficiently capture fine-grained external knowledge, but can serve as an effective guide for leveraging the original retrieved text in RAG.

To answer \textbf{RQ2}, we construct a QA dataset containing up-to-date information beyond the LLM’s internal knowledge. We observe that (1) parametric passage representations can encode new knowledge, with relevant passages exhibiting higher similarity than irrelevant ones; (2) parametric injection induces larger changes in the information flow around the feed-forward networks in deeper layers, suggesting that it primarily provides high-level guidance for answer prediction; and (3) P-RAG alone is substantially weaker than T-RAG at consolidating evidence and reasoning over multi-hop passages, whereas injecting parametric representations into T-RAG enables PT-RAG to outperform T-RAG by a clear margin.

For \textbf{RQ3}, we show that (1) parametric injection significantly improves context faithfulness when retrieved evidence conflicts with the LLM’s internal knowledge, (2) P-RAG is more robust than T-RAG to retrieval noise, and (3) parametric representations trained with passage-grounded QA objectives can transfer to other tasks, such as fact verification and slot filling. 

\begin{table}[!t]
\centering
\caption{Comparison between Token-based RAG, Parametric RAG, and the Combined approach.}
\resizebox{\columnwidth}{!}{
\begin{tabular}{lcccc}
    \toprule
    Dimension & Metric & Token-based RAG & Parametric RAG & Combine Both \\
    \midrule
    \multirow{3}{*}{\textit{Efficiency}} 
    & Inference Latency & High & \textbf{Low} & High \\
    & Training Cost & \textbf{Negligible} & High & High \\
    & Storage Overhead & High & Very High & Very High \\
    \midrule
    
    \multirow{5}{*}{\textit{Capability}} 
    & Information Fidelity & \textbf{Precise} & Lossy & \textbf{Precise} \\
    & Context Faithfulness & Limited & Limited & \textbf{Strong} \\
    & Noise Robustness & Weak & \textbf{Strong} & \textbf{Strong} \\
    & Long Context Support & Constrained & \textbf{Scalable} & \textbf{Scalable} \\
    & Deep Context Integration & Limited & Limited & \textbf{Strong} \\
    \midrule
    
    \multirow{5}{*}{\textit{Usability}}
    & Task Generalization & \textbf{Strong} & Restricted & \textbf{Strong} \\
    & Model Transferability & \textbf{High} & Low & Low \\
    & Interpretability & \textbf{High} & Low & Low \\
    & Update Flexibility & \textbf{Easy} & Hard & Hard \\
    & Ecosystem Maturity & \textbf{Mature} & Nascent & Nascent \\
    \bottomrule
\end{tabular}
}
\label{tab:compare_RAG}
\end{table}

To summarize, in this paper, we present a systematic study of parametric RAG to better understand how parametric knowledge injection behaves in retrieval-augmented generation and when it is most beneficial. Our contributions are four-fold: (1) we reassess P-RAG under a more appropriate evaluation protocol based on HasAnswer accuracy; (2) we introduce a new QA benchmark containing up-to-date knowledge beyond the LLM’s parametric memory to support controlled analysis; (3) we provide in-depth representational and mechanistic analyses of parametric injection, including its semantic encoding properties and its layer-wise effects on model computation; and (4) we evaluate P-RAG under core RAG challenges such as faithfulness under knowledge conflicts, robustness to retrieval noise, and generalization to tasks beyond QA.
Our findings provide practical guidance for future research and development of parametric RAG paradigm.

\section{Related Work}


\subsection{Retrieval-Augmented Generation}
Retrieval-augmented generation (RAG) enhances large language models (LLMs) by retrieving external knowledge to address missing information, improving performance on knowledge-intensive tasks and reducing hallucination~\cite{lewis2020retrieval, zamani2022retrieval, izacard2023atlas}.
Its effectiveness depends on two core challenges: accurate retrieval~\cite{shi2023replug, zhang2025utility, zhang2025distilling} and effective knowledge integration~\cite{zhang2024raft, sun2024redeep}.
Existing efforts on enabling LLMs to better integrate retrieved knowledge can be broadly categorized into three directions:
1) Token-level augmentation~\cite{trivedi2022interleaving, wang2024rat, tang2025injecting}: 
Retrieved documents are inserted directly into the input prompt, often with context refinement.
While simple and widely adopted, it is constrained by the context window limit. Furthermore, as the context length increases, the model suffers from both substantial inference overhead and a degraded capacity to effectively comprehend and utilize the retrieved content.
2) Embedding-level fusion~\cite{izacard2020leveraging, dong2025decoupling, wang2024kblam}: 
Instead of concatenating documents into the prompt, this paradigm encode documents offline and inject the
resulting embeddings into LLMs during inference via cross-attention to mitigate context overhead.
However, the interaction between LLMs and the retrieved content is shallow, often leading to performance degradation—particularly when only a limited number of documents are used.
3) Parametric-level adaptation~\cite{su2025parametric, tan2025dynamic, chen2025privacy, su2025routing}: 
A recent paradigm that converts documents into pluggable model parameters (e.g., LoRA) for direct injection into the LLM. 
\citet{su2025parametric} argues this enables deeper knowledge integration while reducing inference overhead by eliminating document tokens from the prompt.
Our work centers on parametric-level adaptation.
We conduct a systematic analysis of parametric RAG (focusing on P-RAG~\cite{su2025parametric} and DyP-RAG~\cite{tan2025dynamic}), aiming to validate its capability for information preservation and deep interaction.

\subsection{Parametric RAG}

Parametric RAG \cite{su2025parametric, tan2025dynamic, chen2025privacy, su2025routing} is a novel RAG paradigm that avoids inserting documents into the context. 
Instead, it encodes each document into a parametric representation (e.g., LoRA~\cite{hu2022lora}) and injects this representation into the LLM during inference, thereby incorporating external knowledge through parameter updates rather than context augmentation.
To obtain document-specific parameters, P-RAG~\cite{su2025parametric} performs data augmentation for each document, then trains a dedicated LoRA module on this augmented data. 
However, this process requires pre-computing and storing a LoRA for every document in the corpus, leading to significant training and storage overhead. 
To mitigate this, DyP-RAG \cite{tan2025dynamic} introduces a parameter translator that maps documents directly to LoRAs at inference time, achieving comparable performance with substantially reduced cost.
PolyP-RAG~\cite{su2025routing} proposes to encode an entire document collection into a small set of shared latent LoRA experts, which are dynamically routed based on the query.
Yet, existing parametric RAG studies primarily focus on performance gains, often overlooking a fundamental question: do the injected parameters truly encode factual knowledge, or might they merely act as task-specific adapters that improve surface-level formatting?
In contrast, we investigate the underlying mechanisms of parametric injection, examining whether LoRA modules genuinely capture document semantics and enable meaningful knowledge utilization by LLMs.

\section{Preliminary}
\label{section:preliminary}
This section formalizes the inference pipelines of standard RAG and parametric RAG, and subsequently details the parameterization process of two representative approaches: P-RAG and DyP-RAG.

\heading{Standard RAG}
Given a query $q$, a retriever first identifies a set of top-$k$ relevant documents $\mathcal{D} = \{d_1, d_2, ..., d_k\}$ from an external corpus.
In standard RAG, these retrieved documents are explicitly maintained as textual tokens. They are concatenated to form a unified context sequence $c$:
\begin{equation}
    c = \text{Concat}(d_1, d_2, ..., d_k).
\end{equation}
The LLM with parameters $\theta$ then generates the response $y$ by attending to both the explicit context $c$ and the query $q$ within its input window:
\begin{equation}
    y^{\text{RAG}} = \arg\max_{y} P(y \mid c, q; \theta).
\end{equation}

\heading{Parametric RAG}
While effective, standard RAG faces critical challenges, including finite context windows, high inference latency, and degraded performance when processing long contexts.
Parametric RAG addresses these challenges by altering the paradigm of information integration: instead of explicitly inserting documents into the input context, it encodes retrieved documents into model parameters.
During inference, these parametric representations are injected into the LLM, enabling the model to interact with documents at the parameter level without expanding the context length.

Formally, parametric RAG encodes each document $d_i \in \mathcal{D}$ into a parametric representation (e.g., LoRA weights) $\Delta \theta_i = F(d_i)$.
This mapping function $F$ can be realized through diverse approaches, such as offline training (as in P-RAG) or online hypernetwork generation (as in DyP-RAG).
At inference time, the parametric representations of the retrieved top-$k$ documents $\mathcal{D}$ are aggregated to form a unified update. This is typically achieved via linear composition:
\begin{equation}
\Delta \theta_{\mathcal{D}} = \sum_{i=1}^{k} \Delta \theta_i.
\end{equation}
The aggregated parameters are then injected into the LLM. 
The output is generated conditioned only on the query $q$, but with the model parameters adapted to the retrieved content:
\begin{equation}
y^{\text{Para}} = \arg\max_y P(y \mid q; \theta + \Delta \theta_{\mathcal{D}}).
\end{equation}
Parametric RAG can also be seamlessly combined with standard RAG.
In this setting, retrieved documents are included in the input context $c = \text{Concat}(d_1, ..., d_k)$, while the parametric representation is simultaneously injected. The generation process becomes:
\begin{equation}
y^{\text{Combine}} = \arg\max_y P(y \mid c, q; \theta + \Delta \theta_{\mathcal{D}}).
\end{equation}

\heading{Document Parameterization of P-RAG}
P-RAG encodes each document $d_i$ into a parametric representation $\Delta \theta_i$ (implemented as a LoRA module) via gradient-based optimization.
However, as noted in prior work~\cite{allen2023physics}, training solely on raw document text via next-token prediction often fails to internalize factual knowledge effectively.
To address this, P-RAG adopts a data augmentation strategy to enrich the learning signal.

Specifically, for a given document $d_i$, P-RAG prompts the target LLM to generate multiple rewritten variants and a diverse set of QA pairs grounded in the content of $d_i$.
These generated elements are then assembled to construct a document-specific augmented dataset $\mathcal{Z}_i$, where each training sample $z \in \mathcal{Z}_i$ is a sequence formed by concatenating the document, question, and answer.
The LoRA parameters $\Delta \theta_i$ are then optimized by minimizing the negative log-likelihood over these augmented sequences:
\begin{equation}
    \min_{\Delta\theta_i} \sum_{z \in \mathcal{Z}_i} -\log P(z \mid \theta + \Delta\theta_i).
\end{equation}

\heading{Document Parameterization of DyP-RAG}
While P-RAG is effective, the requirement for per-document gradient optimization hinders scalability.
To address this, DyP-RAG~\cite{tan2025dynamic} circumvents the expensive per-document optimization by learning a generalized mapping function $\mathcal{F}_\phi$, implemented as a lightweight hypernetwork.

Formally, given a document $d_i$, DyP-RAG first extracts its semantic representation $h_i$ (e.g., the last hidden state) using the frozen LLM.
The hypernetwork $\mathcal{F}_\phi$ then projects this representation directly into the weight space of the adapter:
\begin{equation}
\Delta \theta_i = \mathcal{F}_\phi(h_i).
\end{equation}
To train $\mathcal{F}_\phi$, DyP-RAG adopts a hybrid alignment strategy, utilizing the offline-optimized parameters from P-RAG as supervision signals.
The optimization objective combines the generative loss on the augmented dataset $\mathcal{Z}_i$ (same as P-RAG) with parameter-level distillation, minimizing both the reconstruction error (MSE) and KL divergence against the target P-RAG parameters.

\begin{table*}[htbp]
\caption{Main experimental results re-evaluated using HasAnswer accuracy (\%). 
Bold numbers indicate the best performance within each model group, underlined results denote the second-best. CWQ denotes ComplexWebQuestions. 
$^{\clubsuit}$ and $^{\diamondsuit}$ indicate a statistically significant difference ($p < 0.05$, two-sided paired t-test) compared to Vanilla LLM and T-RAG, respectively.}
\centering
\resizebox{\textwidth}{!}{
\begin{tabular}{clllllllllll>{\columncolor{lightgray}}l}
    \toprule
    \multicolumn{1}{c}{\multirow{2}{*}{LLM}} & \multicolumn{1}{c}{\multirow{2}{*}{Method}} & \multicolumn{5}{c}{2WikiMultihopQA} & \multicolumn{3}{c}{HotpotQA} & \multicolumn{1}{c}{\multirow{2}{*}{PopQA}} & \multicolumn{1}{c}{\multirow{2}{*}{CWQ}} & \multicolumn{1}{c}{\multirow{2}{*}{Average}} \\ 
    \cmidrule(lr){3-7} \cmidrule(lr){8-10}
    
    \multicolumn{1}{c}{} & \multicolumn{1}{c}{} & Compare & Bridge & Infer & Compose & Total & Bridge & Compare & Total & \multicolumn{1}{c}{} & \multicolumn{1}{c}{} & \multicolumn{1}{c}{} \\ 
    \midrule
    
    \multirow{6}{*}{\begin{tabular}[c]{@{}c@{}}LLaMA3.2- \\ 1B-Instruct\end{tabular}} & Vanilla LLM & \underline{43.00} & \underline{43.00} & 0.66 & 3.66 & 21.00 & 9.00 & 42.33 & 16.00 & 10.66 & 30.33 & 21.96 \\
     & T-RAG & 30.00 & 34.00 & \underline{7.33} & 8.33 & 21.00 & 24.00 & 45.33 & \underline{30.33} & \underline{48.66} & 31.33 & 28.03 \\
     & P-RAG & \textbf{45.00}$^{\diamondsuit}$ & \textbf{43.66}$^{\diamondsuit}$ & 2.00$^{\diamondsuit}$ & 5.00 & \textbf{23.66} & 14.66$^{\clubsuit \diamondsuit}$ & \underline{48.66}$^{\clubsuit}$ & 21.00$^{\clubsuit \diamondsuit}$ & 23.33$^{\clubsuit \diamondsuit}$ & 29.66 & 25.56$^{\clubsuit \diamondsuit}$ \\
     & PT-RAG & 36.33$^{\diamondsuit}$ & 37.33 & \textbf{7.66}$^{\clubsuit}$ & \textbf{9.66}$^{\clubsuit}$ & \underline{22.00} & \underline{25.33}$^{\clubsuit}$ & \textbf{51.33}$^{\clubsuit \diamondsuit}$ & \textbf{31.00}$^{\clubsuit}$ & \underline{48.66}$^{\clubsuit}$  & \textbf{36.66}$^{\clubsuit \diamondsuit}$ & \textbf{30.60}$^{\clubsuit \diamondsuit}$ \\
     & DyP-RAG & 42.66$^{\diamondsuit}$ & 33.00$^{\clubsuit}$ & 1.33$^{\diamondsuit}$ & 4.00$^{\diamondsuit}$ & 21.00 & 11.00$^{\diamondsuit}$ & 44.00 & 14.66$^{\diamondsuit}$ & 12.33$^{\diamondsuit}$ & 31.66 & 21.56$^{\diamondsuit}$ \\
     & DyPT-RAG & 35.66$^{\clubsuit \diamondsuit}$ & 31.66$^{\clubsuit}$ & 7.00$^{\clubsuit}$ & \underline{8.66}$^{\clubsuit}$ & 21.66 & \textbf{25.66}$^{\clubsuit}$ & 44.00 & 28.00$^{\clubsuit}$ & \textbf{49.00}$^{\clubsuit}$  & \underline{34.00} & \underline{28.53}$^{\clubsuit}$ \\
     
     \midrule
     
    \multirow{6}{*}{\begin{tabular}[c]{@{}c@{}}Qwen2.5- \\ 1.5B-Instruct\end{tabular}} & Vanilla LLM & \textbf{28.33} & 29.33 & 0.66 & 5.00 & 14.00 & 7.00 & 41.00 & 13.33 & 13.00 & 27.66 & 17.93 \\
     & T-RAG & 24.33 & 21.66 & 5.66 & 4.00 & 14.66 & \underline{23.33} & 46.66 & \underline{27.33} & \textbf{50.00} & 23.33 & 24.10 \\
     & P-RAG & 27.66 & \textbf{34.66}$^{\clubsuit \diamondsuit}$ & 2.66$^{\clubsuit}$ & 5.00 & \textbf{16.33} & 10.00$^{\diamondsuit}$ & 38.00$^{\diamondsuit}$ & 14.33$^{\diamondsuit}$ & 21.33$^{\clubsuit \diamondsuit}$ & \underline{31.66}$^{\diamondsuit}$ & 20.16$^{\clubsuit \diamondsuit}$ \\
     & PT-RAG & \underline{28.00} & 27.33$^{\diamondsuit}$ & \textbf{7.66}$^{\clubsuit}$ & \textbf{8.00}$^{\diamondsuit}$ & \underline{15.66} & \textbf{25.33}$^{\clubsuit}$ & \textbf{52.33}$^{\clubsuit \diamondsuit}$ & \textbf{31.66}$^{\clubsuit \diamondsuit}$ & 46.00$^{\clubsuit}$ & 28.99$^{\diamondsuit}$ & \textbf{27.10}$^{\clubsuit \diamondsuit}$ \\ 
    & DyP-RAG & 23.33 & \underline{34.33}$^{\diamondsuit}$ & 0.66$^{\diamondsuit}$ & 5.00 & 14.66 & 7.33$^{\diamondsuit}$ & 39.66$^{\diamondsuit}$ & 13.00$^{\diamondsuit}$ & 14.00$^{\diamondsuit}$ & \textbf{33.33}$^{\clubsuit \diamondsuit}$ & 18.53$^{\diamondsuit}$ \\
     & DyPT-RAG & 25.33 & 25.00 & \underline{6.33}$^{\clubsuit}$ & \underline{7.66}$^{\diamondsuit}$ & 13.66 & 22.33$^{\clubsuit}$ & \underline{48.66}$^{\clubsuit}$ & 26.66$^{\clubsuit}$ & \underline{47.00}$^{\clubsuit}$ & 31.33$^{\diamondsuit}$ & \underline{25.40}$^{\clubsuit \diamondsuit}$ \\     
     \midrule
     
    \multirow{4}{*}{\begin{tabular}[c]{@{}c@{}}Qwen2.5- \\ 7B-Instruct\end{tabular}} & Vanilla LLM & \underline{49.66} & \underline{47.66} & 1.66 & 7.00 & \underline{25.00} & 14.00 & \underline{61.66} & 20.33 & 18.00 & 33.66 & 27.86 \\
     & T-RAG & 45.00 & 41.33 & \underline{10.66} & 8.00 & 22.66 & \underline{31.66} & 54.65 & \underline{34.66} & \underline{36.66} & 26.00 & 31.33 \\
     & P-RAG & \textbf{56.33}$^{\clubsuit \diamondsuit}$ & \textbf{49.00}$^{\diamondsuit}$ & 2.33$^{\diamondsuit}$ & \underline{10.66}$^{\clubsuit}$ & \textbf{28.66}$^{\diamondsuit}$ & 20.00$^{\clubsuit \diamondsuit}$ & \textbf{63.00}$^{\diamondsuit}$ & 26.66$^{\clubsuit \diamondsuit}$ & 31.33$^{\clubsuit}$ & \textbf{44.66}$^{\clubsuit \diamondsuit}$ & \underline{33.26}$^{\clubsuit \diamondsuit}$ \\
     & PT-RAG & 46.66 & 37.33$^{\clubsuit}$ & \textbf{12.00}$^{\clubsuit}$ & \textbf{11.33}$^{\clubsuit \diamondsuit}$ & \underline{25.00} & \textbf{35.33}$^{\clubsuit}$ & 57.66 & \textbf{38.66}$^{\clubsuit}$ & \textbf{43.33}$^{\clubsuit \diamondsuit}$ & \underline{37.00}$^{\diamondsuit}$ & \textbf{34.43}$^{\clubsuit \diamondsuit}$ \\ 

     \bottomrule
\end{tabular}
}
\label{tab:reproduction_results_llm}
\end{table*}

\section{Reassessing Parametric RAG}
This section investigates \textbf{RQ1}: \textit{Do P-RAG methods still outperform T-RAG when evaluated with answer-presence accuracy, rather than length-sensitive F1?}
%
We conduct a rigorous re-evaluation of P-RAG and DyP-RAG using HasAnswer accuracy.
By decoupling correctness from response verbosity, we aim to uncover the true performance gap between parametric injection and standard T-RAG.


\subsection{Experimental Setup}
\label{sec:repro_setup}
Our experimental framework largely aligns with established configurations in prior studies~\cite{su2025parametric, tan2025dynamic} regarding datasets and baselines, but incorporates targeted adjustments to a few problematic settings to ensure a fairer and more interpretable evaluation.

\heading{Evaluation Metric}
Prior studies on parametric RAG predominantly relied on the F1 score.
We argue that this metric is unreliable for assessing knowledge injection, as it is highly sensitive to response length and lexical overlaps, often failing to capture semantic equivalence.
As illustrated in Figure~\ref{fig:case_study}, T-RAG provides a correct but detailed answer, yielding in a low F1 of 0.33 due to the length penalty.
In contrast, P-RAG generates a concise output, achieving a perfect F1 of 1.0.
%
Consequently, F1 exhibits a bias that inflates the performance of parametric RAG—rewarding its tendency toward template-like brevity rather than genuine knowledge acquisition.

To mitigate these limitations, we adopt \textbf{HasAnswer accuracy} as our primary evaluation metric.
This metric checks whether the generated response contains the ground-truth answer or a consistent equivalent, decoupling correctness from verbosity.
We implement this using an LLM-as-a-judge framework, drawing on recent work demonstrating its superior reliability~\cite{ho2025llm}.
Specifically, we employ Qwen2.5-32B-Instruct~\cite{qwen2.5} as the judge, prompting it to determine whether the model response is semantically equivalent the ground truth.
To validate the reliability of this approach, we manually annotated 300 randomly sampled instances and observed a 96.7\% agreement rate between the LLM judge and human evaluators, confirming its robustness.


\heading{Inference and Parameterization Setup}
To strictly evaluate the effectiveness of knowledge injection, we enforce a zero-shot setting across all experiments.
While the original P-RAG setup utilized few-shot prompts during both parameterization and inference on specific datasets.
We argue that this practice risks conflating task-specific patterns with the factual content of documents.
Therefore, we remove few-shot examples from the inference prompts for all methods, and likewise exclude them from P-RAG's training data.
To ensure reproducibility, all generations utilize greedy decoding.

Regarding the parameterization process, both P-RAG and DyP-RAG are configured to apply LoRA to the FFN layers with rank $r=2$ and scaling factor $\alpha=32$.
Specific implementation details are as follows:
(i) P-RAG: We strictly follow the pipeline described in Section \ref{section:preliminary}. Consistent with the original work~\cite{su2025parametric}, each document is paired with one rewritten variant and three QA pairs for training. LoRA modules are optimized for one epoch with a learning rate of $3 \times 10^{-4}$.
(ii) DyP-RAG: We utilize the official pre-trained hypernetwork checkpoints provided by the authors~\cite{tan2025dynamic}, which generate LoRA adapters adhering to the aforementioned configuration.

\heading{Datasets}
We adopt the same four datasets as prior studies~\cite{su2025parametric,tan2025dynamic}: 2WikiMultihopQA~\cite{ho2020constructing}, HotpotQA~\cite{yang2018hotpotqa}, ComplexWebQuestions~\cite{talmor2018web}, and PopQA~\cite{mallen2023not}.
Consistent with prior work, we evaluate on the first 300 questions for each dataset. 
Specifically for 2WikiMultihopQA and HotpotQA, the ``Total'' column in Table~\ref{tab:reproduction_results_llm} reports results on the first 300 questions of the entire dataset, while each sub-task column shows results on the first 300 questions within that sub-task.
For each question, we utilize BM25 to retrieval the top-3 relevant passages from a Wikipedia dump~\cite{karpukhin2020dense}.

\heading{Methods and Models}
We evaluate six methods, covering standard token-based, parametric, and hybrid paradigms:
\begin{itemize}[leftmargin=*,itemsep=0pt,topsep=2pt,parsep=0pt]
    \item \textbf{\Vanilla:} The base model generating responses directly without retrieval.
    \item \textbf{\RAG:} The standard token-based RAG approach where retrieved documents are explicitly appended to the input context.
    \item \textbf{\PRAG:} The static parametric method where retrieved documents are encoded via offline-trained LoRA adapters.
    \item \textbf{\PRAGCombine:} A hybrid variant combining \PRAG and \RAG, utilizing both parametric injection and explicit textual context.
    \item \textbf{\DyPRAG:} The dynamic parametric method where adapters are generated on-the-fly by a hypernetwork.
    \item \textbf{\DyPRAGCombine:} The hybrid counterpart of \DyPRAG, combining dynamic parameter generation with token-based context.
\end{itemize}

Experiments are conducted on three open-source LLMs spanning different families and scales: LLaMA3.2-1B-Instruct~\cite{grattafiori2024llama}, Qwen2.5-1.5B-Instruct, and Qwen2.5-7B-Instruct~\cite{qwen2.5}.
Note that for Qwen2.5-7B-Instruct, we exclude \DyPRAG and \DyPRAGCombine from evaluation, as official checkpoints are not available for this model.

\subsection{Experimental Results}
\label{sec:repro_results}

Table~\ref{tab:reproduction_results_llm} presents the re-evaluation results based on semantic accuracy.
Under this rigorous metric, we observe three critical findings that offer a revised perspective on the efficacy of parametric knowledge injection:
\textbf{(1) Hybrid paradigms consistently achieve superior performance.} 
Aligning with prior findings~\cite{su2025parametric, tan2025dynamic}, methods that combine parametric injection with explicit textual context (i.e., \PRAGCombine and \DyPRAGCombine) yield the best results across almost all datasets and models. 
\textbf{(2) \PRAG improves over \Vanilla but falls short of \RAG.} 
Contrary to prior findings, our results show that while \PRAG significantly outperforms the \Vanilla baseline, it does not consistently surpass \RAG in most settings. 
We attribute this discrepancy to the metric bias discussed in Section~\ref{sec:repro_setup}. Previous F1-based evaluations likely overestimated \PRAG's effectiveness by rewarding its tendency to generate concise, template-like responses.
Our semantic evaluation eliminates this inflation, providing a faithful assessment that exposes the true capability of parametric injection.
\textbf{(3) \DyPRAG fails to demonstrate meaningful gains.} 
Diverging significantly from the original report, \DyPRAG shows only marginal improvement over \Vanilla under this evaluation, with no statistically significant difference observed in average scores.
This suggests that, despite its architectural novelty, the parameters dynamically generated by the hypernetwork struggle to encode semantically robust knowledge—lagging behind both direct parameter optimization (\PRAG) and explicit context (\RAG).

\section{Analyzing Parametric Knowledge Encoding}
\label{section:analyzing}

This section addresses our \textbf{RQ2}: \textit{Do parametric representations encode passage semantics effectively, and how does parametric injection affect the LLM's computation?}
While parametric injection introduces a novel pathway for knowledge integration, the nature of the encoded information remains opaque.
To shed light on these mechanisms, we conduct a systematic investigation spanning three dimensions:
(1) assessing knowledge fidelity using a controlled new-knowledge benchmark;
(2) probing the semantic specificity and layer-wise impact of the injected parameters; and
(3) evaluating their capability to support complex, multi-hop reasoning.


\subsection{Experimental Setup}
\label{sec:news-setup}

\heading{Dataset Construction}
To accurately assess the fidelity of parametric knowledge encoding, it is essential to eliminate the interference of the model's pre-existing internal memory.
To this end, we construct the \textbf{\Dataset} dataset comprising information entirely absent from the LLMs' training corpora.
This ensures a controlled evaluation where the model must rely exclusively on the provided external information to answer correctly.

Specifically, we collected 300 news articles from major outlets (e.g., BBC News\footnote{\url{https://www.bbc.com/news}}) published in 2025.
These articles appeared after the knowledge cutoff dates of all LLMs used in this work.
We employ Qwen2.5-32B-Instruct to generate two types of QA pairs grounded in each article:
(i) Factual QA: questions targeting specific facts within the content;
(ii) Multihop QA: questions requiring the synthesis of multiple facts from the article.
This process yielded a total of 600 QA pairs.
Finally, we segmented the articles into non-overlapping chunks based on sentence boundaries, with passage lengths matching the granularity of passages in Section~\ref{sec:repro_setup}.
This processing resulted in a total of 807 passages, which serve as the direct source for knowledge injection.

\heading{Inference and Parameterization}
During inference, we directly provide the model with the question and the ground-truth passages, bypassing the retrieval stage to exclude the interference of retrieval quality.
Regarding parameterization:
For \PRAG, we train a LoRA adapter for each passage using the same hyperparameters described in Section~\ref{sec:repro_setup}.
For \DyPRAG, given its claimed strong generalizability, we employ the same official hypernetwork checkpoint used in Section~\ref{sec:repro_setup} to directly generate parameters.
All other configurations—including parameterization pipeline and evaluation metric—remain consistent with the main experiments.

\begin{table}[tbp]
\centering
\caption{Performance comparison on \Dataset dataset, where the model must rely on external information to answer correctly.
Symbols $^{\clubsuit}$ and $^{\diamondsuit}$ follow the definitions in Table~\ref{tab:reproduction_results_llm}.}
\resizebox{\columnwidth}{!}{
\begin{tabular}{clll>{\columncolor{lightgray}}l}
    \toprule

    \multicolumn{1}{c}{\multirow{2}{*}{LLM}} & \multicolumn{1}{c}{\multirow{2}{*}{Method}} & \multicolumn{3}{c}{\Dataset} \\ 
    \cmidrule(lr){3-5}
    
    \multicolumn{1}{c}{} & \multicolumn{1}{c}{} & Factual QA & Multihop QA & \multicolumn{1}{l}{\multirow{1}{*}{Average}} \\ 
    \midrule
    
    \multirow{6}{*}{\begin{tabular}[c]{@{}c@{}}LLaMA3.2- \\ 1B-Instruct\end{tabular}} & \Vanilla & 9.66 & 7.33 & 8.50 \\
     & \RAG & 71.66 & 56.66 & 64.16 \\
     & \PRAG & 13.00$^{\clubsuit \diamondsuit}$ & 14.00$^{\clubsuit \diamondsuit}$ & 13.50$^{\clubsuit \diamondsuit}$ \\
     & \PRAGCombine & \textbf{78.33}$^{\clubsuit \diamondsuit}$ & \textbf{64.00}$^{\clubsuit \diamondsuit}$  & \textbf{71.16}$^{\clubsuit \diamondsuit}$ \\
     & \DyPRAG & 10.00$^{\diamondsuit}$ & 8.33$^{\diamondsuit}$ & 9.16$^{\diamondsuit}$\\
     & \DyPRAGCombine & 75.00$^{\clubsuit}$ & 57.66$^{\clubsuit}$  & 66.33$^{\clubsuit}$ \\
     \midrule

    \multirow{6}{*}{\begin{tabular}[c]{@{}c@{}}Qwen2.5- \\ 1.5B-Instruct\end{tabular}} & \Vanilla & 5.66 & 8.66 & 7.16 \\
     & \RAG & 70.00 & 57.99 & 64.00 \\
     & \PRAG & 7.33$^{\diamondsuit}$ & 11.00$^{\diamondsuit}$ & 9.16$^{\clubsuit \diamondsuit}$ \\
     & \PRAGCombine & \textbf{75.66}$^{\clubsuit \diamondsuit}$ & \textbf{65.00}$^{\clubsuit \diamondsuit}$  & \textbf{70.33}$^{\clubsuit \diamondsuit}$ \\
     & \DyPRAG & 3.33$^{\clubsuit \diamondsuit}$ & 8.66$^{\diamondsuit}$ & 6.00$^{\diamondsuit}$ \\
     & \DyPRAGCombine & 75.33$^{\clubsuit \diamondsuit}$ & 63.66$^{\clubsuit \diamondsuit}$  & 69.5$^{\clubsuit \diamondsuit}$ \\
     \midrule

    \multirow{4}{*}{\begin{tabular}[c]{@{}c@{}}Qwen2.5- \\ 7B-Instruct\end{tabular}} & \Vanilla & 13.66 & 18.00 & 15.83 \\
     & \RAG & 74.00 & 65.66 & 69.83 \\
     & \PRAG & 21.66$^{\clubsuit \diamondsuit}$ & 19.00$^{\diamondsuit}$ & 20.33$^{\clubsuit \diamondsuit}$ \\
     & \PRAGCombine & \textbf{80.00}$^{\clubsuit \diamondsuit}$ & \textbf{74.33}$^{\clubsuit \diamondsuit}$  & \textbf{77.16}$^{\clubsuit \diamondsuit}$ \\
     
    \bottomrule
\end{tabular}
}
\label{tab:news-result}
\end{table}


\subsection{Evaluation on New-Knowledge QA}
\label{sec:news-result}

\begin{figure}[t]
    \centering
    \includegraphics[width=\linewidth]{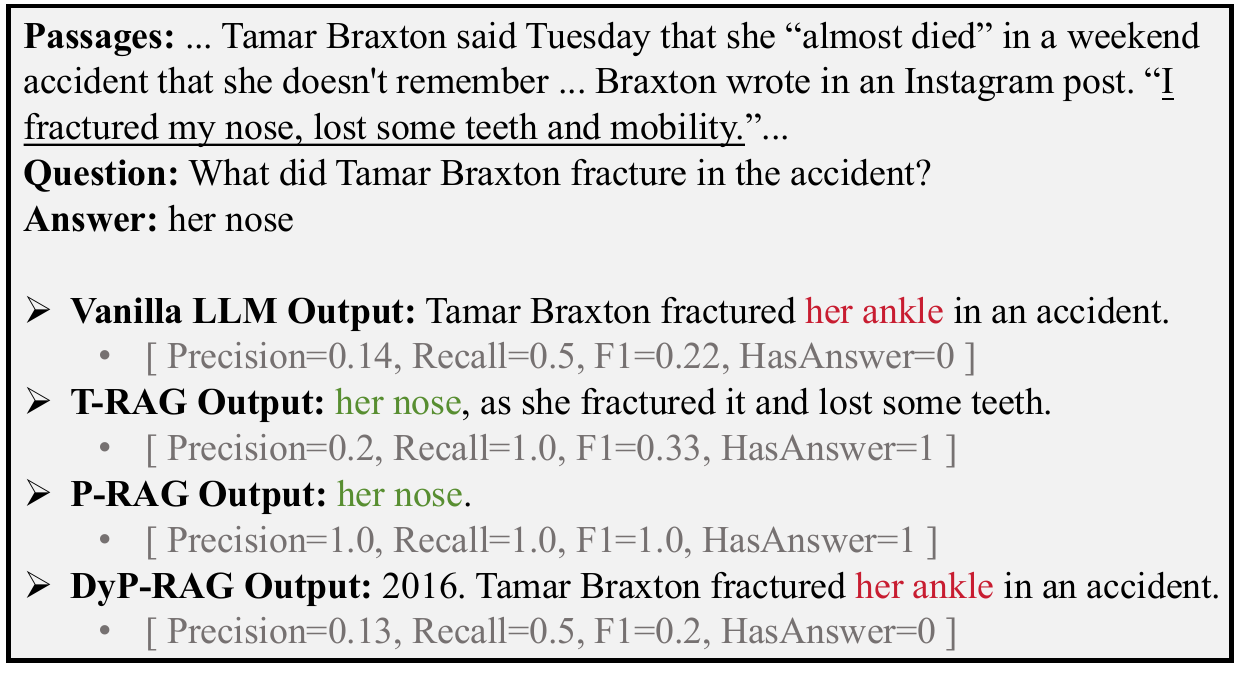}
    \caption{An example from the \Dataset dataset involving an event that occurred after the LLM's training cutoff. }
    \label{fig:case_study}
\end{figure}

Table~\ref{tab:news-result} presents the performance on the \Dataset dataset, a controlled setting designed to exclude reliance on the model's internal (pre-trained) knowledge.
The results reveal four key findings:
\textbf{(1) \PRAG successfully encodes factual knowledge.}
\PRAG significantly outperforms the \Vanilla baseline across most settings.
Qualitatively, the case study in Figure~\ref{fig:case_study} provides direct evidence: when queried about an event outside its training horizon, \Vanilla hallucinates, whereas \PRAG produces the correct answer, indicating that factual knowledge is stored in the parameters.
\textbf{(2) \DyPRAG fails to inject novel information.}
\DyPRAG performs poorly in this strict setting, showing no statistically significant difference from \Vanilla.
This suggests that the dynamically generated parameters do not genuinely capture the factual content of documents, contradicting the original claims regarding its effectiveness and generalization capabilities.
\textbf{(3) Parametric encoding suffers from limited encoding fidelity.}
Both \PRAG and \DyPRAG exhibit a substantial performance gap compared to token-based \RAG.
This indicates that current parametric representations struggle to preserve fine-grained details from source documents, leading to significant information loss compared to explicit textual context.
\textbf{(4) Hybrid paradigms consistently achieve the best performance.}
Both \PRAGCombine and \DyPRAGCombine yield the highest accuracy, significantly surpassing \RAG.
This suggests that even if standalone parametric encoding is imperfect, it provides a valuable auxiliary signal that complements the explicit context to enhance the final generation.

\begin{figure*}[htbp]
    \centering
    \includegraphics[width=\linewidth]{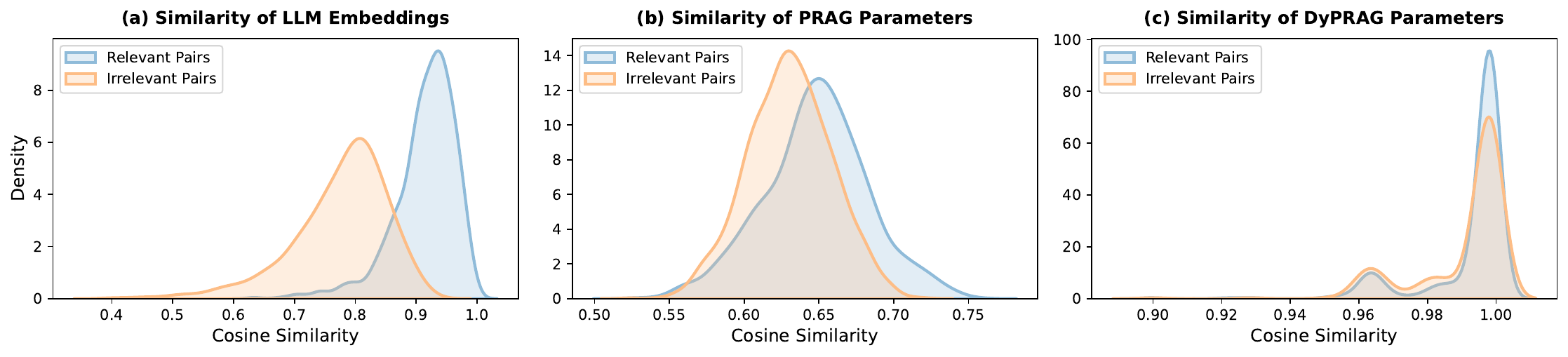}
    \caption{
    Cosine similarity distributions for relevant versus irrelevant passage pairs on Qwen2.5-1.5B-Instruct, measured using three distinct representations:
    (a) last-token hidden state embeddings from from the LLM;
    (b) flattened LoRA parameters (\texttt{up\_proj}) from \PRAG; and
    (c) flattened LoRA parameters (\texttt{up\_proj}) from \DyPRAG.
    }
    \label{fig:sim_distribution}
\end{figure*}

\begin{figure*}[htbp]
    \centering
    \includegraphics[width=0.95\linewidth]{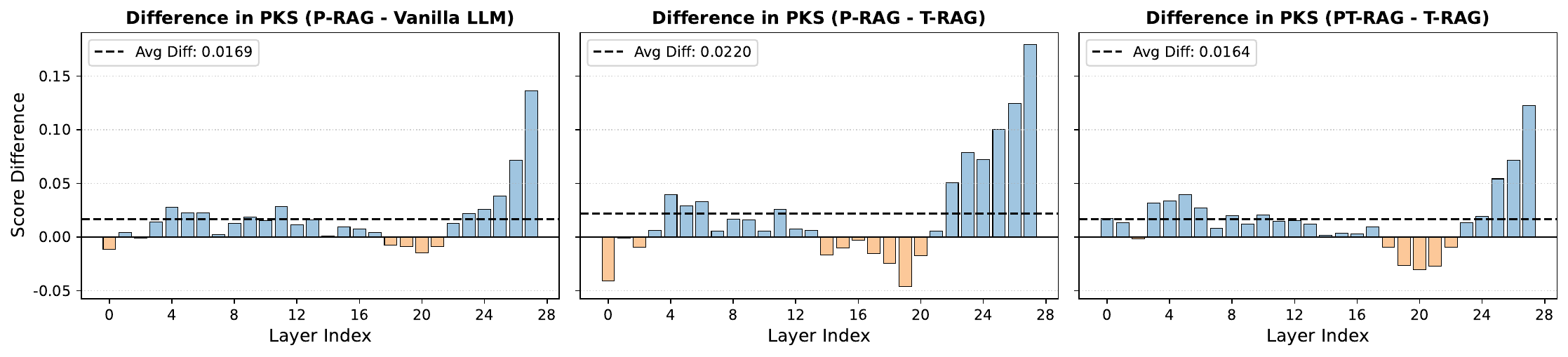}
    \caption{Layer-wise PKS differences on Qwen2.5-1.5B-Instruct.
    Values represent the target method's PKS minus the baseline's (e.g., \PRAG \ - \Vanilla).
    Positive bars indicate that the target method injects more parametric knowledge at that specific layer relative to the baseline.
    The horizontal dashed line denotes the average difference across all layers.}
    \label{fig:pks_compare}
\end{figure*}

\subsection{Specificity of Parametric Representations}
\label{sec:specificity}

We further investigate the encoding mechanism by examining the specificity of the parametric representations.
The intuition is straightforward: if parameters faithfully encode document-specific knowledge,
they should exhibit specificity—that is, parameters derived from semantically related content should be close in vector space, while those from unrelated content should be distinct.

We quantify this by comparing the similarity distributions of two distinct groups constructed from the \Dataset dataset:
(i) Relevant Pairs: Passages segmented from the same source article. Since they share thematic context and factual grounding, their representations are expected to be similar.
(ii) Irrelevant Pairs: Passages derived from different articles. Given their distinct factual content, their representations are expected to be divergent.
For \PRAG and \DyPRAG, we calculate the cosine similarity between the flattened LoRA weight matrices for each corresponding layer and report the average similarity across all layers.
Due to space constraints, we focus on the \texttt{up\_proj} module of Qwen2.5-1.5B-Instruct as a representative example (other models and FFN modules exhibit consistent trends).
To provide a reference baseline for semantic distinguishability, we also compute the similarity of the LLM embeddings for the same pairs, where the embedding is defined as the last-token hidden state of the LLM given the passage input.

Figure~\ref{fig:sim_distribution} illustrates the similarity distributions.
The contrast between these distributions provides a mechanistic explanation for the performance gaps observed earlier:
\textbf{(1) \PRAG captures semantic content but with limited fidelity.}
As shown in Figure~\ref{fig:sim_distribution}(b), \PRAG exhibits measurable specificity: relevant pairs yield higher average similarity than irrelevant ones, confirming that its parameters do encode document-specific semantic information.
However, compared to the clear separation observed in the LLM embeddings (Figure~\ref{fig:sim_distribution}(a)), the parameter distributions of \PRAG heavily overlap.
This indicates that while semantic signals are preserved at a coarse level, fine-grained distinctions are lost.
\textbf{(2) \DyPRAG may suffer from hypernetwork collapse.}
In Figure~\ref{fig:sim_distribution}(c), the distributions for relevant and irrelevant pairs are virtually indistinguishable for \DyPRAG, with scores for all pairs saturating at an extremely high range.
Combining this observation with the results in Sections~\ref{sec:repro_results} and \ref{sec:news-result}, where \DyPRAG showed no significant difference over \Vanilla, we infer that the hypernetwork fails to effectively condition on the input content.
Instead, it appears to generate generic, task-specific patterns regardless of the specific document.

\subsection{Layer-wise Effects of Parametric Injection}
\label{sec:pks}

To understand where and how parametric injection influences the generation process, we analyze the model's internal dynamics using the \textbf{Parametric Knowledge Score (PKS)}~\cite{sun2024redeep}.
Building on the insight that FFNs function as key-value memories for factual knowledge~\cite{geva2021transformer}, PKS quantifies the information contribution of each layer based on the intuition that \textit{if an FFN injects critical knowledge, the token distribution should shift significantly after passing through it.}
\looseness=-1

Formally, for each generated token at position $n$ and layer $l$, PKS calculates the Jensen–Shannon divergence (JSD) between the vocabulary distributions before and after the FFN block:
\begin{equation}
    P_n^l = \text{JSD}\left(q(\mathbf{h}_{n,l}^{\text{in}}) \parallel q(\mathbf{h}_{n,l}^{\text{out}})\right),
\end{equation}
where $\mathbf{h}_{n,l}^{\text{in}}$ and $\mathbf{h}_{n,l}^{\text{out}}$ denote the hidden states entering and exiting the FFN block, respectively.
The distribution $q(\mathbf{h})$ is defined as:
\begin{equation}
    q(\mathbf{h}) = \text{softmax}(\text{LayerNorm}(\mathbf{h}) W_U),
\end{equation}
where $W_U$ is the unembedding matrix.
A higher PKS value (i.e., a larger JSD) indicates a more substantial shift in the predictive distribution, suggesting greater incorporation of parametric knowledge at that layer.
The final PKS for layer $l$ is obtained by averaging $P_n^l$ over all tokens in the generated response.

We conduct this analysis on the \Dataset dataset, focusing on \PRAG using Qwen2.5-1.5B-Instruct and excluding \DyPRAG given the representation collapse observed in Section~\ref{sec:specificity}.
To isolate the specific contribution of parametric injection, Figure~\ref{fig:pks_compare} visualizes the per-layer difference in PKS for three comparative pairs:
(a) \PRAG minus \Vanilla;
(b) \PRAG minus \RAG; and
(c) \PRAGCombine minus \RAG.
Two key patterns emerge:
\textbf{(1) Parametric injection increases parametric knowledge contribution.}
As indicated by the dashed lines in Figure~\ref{fig:pks_compare}, the average PKS difference is consistently positive across all comparisons.
This suggests that parametric injection introduces additional parametric knowledge into the generation flow, resulting in a greater information contribution from the FFN layers compared to the baselines.
\textbf{(2) The effect is concentrated in deep layers.}
The PKS gains are not uniformly distributed but are sharply concentrated in the late layers.
This distribution suggests that parametric injection primarily influences the high-level semantic integration stage.
Rather than encoding low-level lexical features, the parameters likely store abstract document priors—high-level semantics—that help steer the final generation toward the correct context.


\subsection{Impact on Multi-hop Reasoning}
\label{sec:gold}

\begin{table}[tbp]
\centering
\caption{
    Performance on 2WikiMultihopQA and HotpotQA using gold passages.
    The ``With QA-LoRA'' columns introduce a generic task-adaptation LoRA module to isolate the gains attributable to specific document knowledge encoding.
    Symbols $^{\clubsuit}$ and $^{\diamondsuit}$ follow the definitions in Table~\ref{tab:reproduction_results_llm}.}
\resizebox{\columnwidth}{!}{
\begin{tabular}{clllll}
    \toprule
    \multicolumn{1}{c}{\multirow{2}{*}{LLM}} & \multicolumn{1}{c}{\multirow{2}{*}{Method}} & \multicolumn{2}{c}{Without QA-LoRA} & \multicolumn{2}{c}{With QA-LoRA}  \\ 
    \cmidrule(lr){3-4} \cmidrule(lr){5-6}
    
    \multicolumn{1}{c}{} & \multicolumn{1}{c}{} & 2Wiki & HotpotQA & 2Wiki & HotpotQA \\ 
    \midrule
    
    \multirow{4}{*}{\begin{tabular}[c]{@{}c@{}}LLaMA3.2- \\ 1B-Instruct\end{tabular}} & \Vanilla & 21.00 & 16.00 & 20.66 & 16.66 \\
     & \RAG & \underline{39.33} & \underline{63.66} & \underline{45.66} & \underline{63.66} \\
     & \PRAG & 23.33$^{\diamondsuit}$ & 21.66$^{\clubsuit \diamondsuit}$ & 23.66$^{\diamondsuit}$ & 22.33$^{\clubsuit \diamondsuit}$ \\
     & \PRAGCombine & \textbf{46.00}$^{\clubsuit \diamondsuit}$ & \textbf{69.33}$^{\clubsuit \diamondsuit}$ & \textbf{46.33}$^{\clubsuit}$ & \textbf{69.33}$^{\clubsuit \diamondsuit}$ \\
     \midrule

    \multirow{4}{*}{\begin{tabular}[c]{@{}c@{}}Qwen2.5- \\ 1.5B-Instruct\end{tabular}} & \Vanilla & 14.00 & 13.33 & 24.66 & 15.66 \\
     & \RAG & \underline{28.33} & \underline{53.66} & \textbf{47.33} & \underline{63.00} \\
     & \PRAG & 18.33$^{\clubsuit \diamondsuit}$ & 15.00$^{\diamondsuit}$ & 18.66$^{\clubsuit \diamondsuit}$ & 16.00$^{\diamondsuit}$ \\
     & \PRAGCombine & \textbf{40.00}$^{\clubsuit \diamondsuit}$ & \textbf{67.00}$^{\clubsuit \diamondsuit}$ & \underline{40.00}$^{\clubsuit \diamondsuit}$ & \textbf{67.33}$^{\clubsuit}$ \\
     \midrule
     
    \multirow{4}{*}{\begin{tabular}[c]{@{}c@{}}Qwen2.5- \\ 7B-Instruct\end{tabular}} & \Vanilla & 25.00 & 20.33 & 28.00 & 20.33 \\
     & \RAG & \underline{56.99} & \underline{67.33} & \underline{54.33} & \underline{73.00} \\
     & \PRAG & 30.66$^{\clubsuit \diamondsuit}$ & 32.33$^{\clubsuit \diamondsuit}$ & 30.66$^{\diamondsuit}$ & 32.66$^{\clubsuit \diamondsuit}$ \\
     & \PRAGCombine & \textbf{64.00}$^{\clubsuit \diamondsuit}$ & \textbf{80.66}$^{\clubsuit \diamondsuit}$ & \textbf{64.00}$^{\clubsuit \diamondsuit}$ & \textbf{80.33}$^{\clubsuit \diamondsuit}$ \\
    \bottomrule
\end{tabular}
}
\label{tab:gold_results}
\end{table}

By injecting external document knowledge directly into the model's FFNs via parameters, parametric RAG aims to facilitate a deeper level of information interaction compared to the surface-level attention mechanism.
This FFN-based knowledge infusion may provide a more direct pathway for evidence fusion—potentially benefiting tasks that require multi-step reasoning and evidence consolidation.
In this section, we empirically investigate this capability.

We conduct experiments on two multi-hop QA datasets: HotpotQA~\cite{yang2018hotpotqa} and 2WikiMultihopQA~\cite{ho2020constructing}, using the first 300 examples from each dataset.
To isolate the model's intrinsic reasoning and integration capabilities—and eliminate confounding effects from retrieval errors—we use the gold passages for both parametric encoding and explicit context.
%
Given the weak conditioning on document content observed for \DyPRAG in Section~\ref{sec:specificity}, we exclude it from this evaluation.
However, since \PRAG is also trained using a QA-style objective, its performance gains could stem not only from document-specific knowledge but also from general task adaptation.
To disentangle these effects, we introduce an additional baseline: \textbf{QA-LoRA}.
Specifically, we sample 400 instances from the same data-augmented corpus used to train \PRAG and train a single, general-purpose LoRA module under identical settings.
This allows us to compare \PRAG (equipped with document-specific parameters) against Vanilla LLM + QA-LoRA (equipped with general task parameters), thereby quantifying the benefits of parametric knowledge injection.

Table~\ref{tab:gold_results} presents the results on the multi-hop reasoning datasets using gold passages.
We observe three critical patterns:
\textbf{(1) Standalone parameters struggle with multi-hop reasoning.}
\PRAG alone significantly underperforms \RAG in these settings.
This confirms that due to the information loss in parametric encoding, \PRAG alone fails to retain the fine-grained details necessary for precise multi-step reasoning compared to explicit context.
\textbf{(2) Parametric injection facilitates multi-step reasoning when combined with explicit context.}
\PRAGCombine consistently achieves the best performance.
Notably, the performance gap between \PRAGCombine and \RAG is substantially larger here than results in Table~\ref{tab:reproduction_results_llm}.
This aligns with our analysis in Section~\ref{sec:pks}: the injected parameters provide high-level semantic guidance, which helps the model better locate and integrate relevant information from the explicit context.
\textbf{(3) Gains stem primarily from document knowledge.}
Adding the QA-LoRA to \PRAG or \PRAGCombine results in negligible performance changes, indicating that \PRAG already adapts well to the task format.
Crucially, even when \Vanilla and \RAG are equipped with QA-LoRA, they still fail short of \PRAGCombine.
This suggests that the performance gains stem primarily from the document-specific knowledge encoded in the parameters, rather than from generic task-adaptation effects.


\section{P-RAG under Critical RAG Challenges}
\label{section:robustness}

This section addresses \textbf{RQ3}: \textit{How does P-RAG behave under core RAG challenges, including faithfulness under knowledge conflicts, robustness to retrieval noise, and generalization to tasks beyond QA?}
Moving beyond mechanistic analysis, we evaluate the practical utility of the parametric paradigm by testing it against three critical challenges in RAG:
(1) maintaining faithfulness when external evidence conflicts with internal priors;
(2) exhibiting robustness against noisy retrieval results; and
(3) demonstrating generalization to downstream tasks.
%


\subsection{Experimental Setup}
Building on the findings in Section~\ref{section:analyzing}, which suggest that \DyPRAG struggles to encode document-specific semantics, we restrict our analysis in this section to \PRAG.
%
Unless otherwise specified, all experiments follow the protocols established in Section~\ref{sec:repro_setup}, with task-specific adaptations described below. 
In all settings, the same passages are used both for parametric encoding (i.e., training the LoRA parameters) and as explicit context during inference.

\begin{figure*}[htbp]
    \centering
    \includegraphics[width=\linewidth]{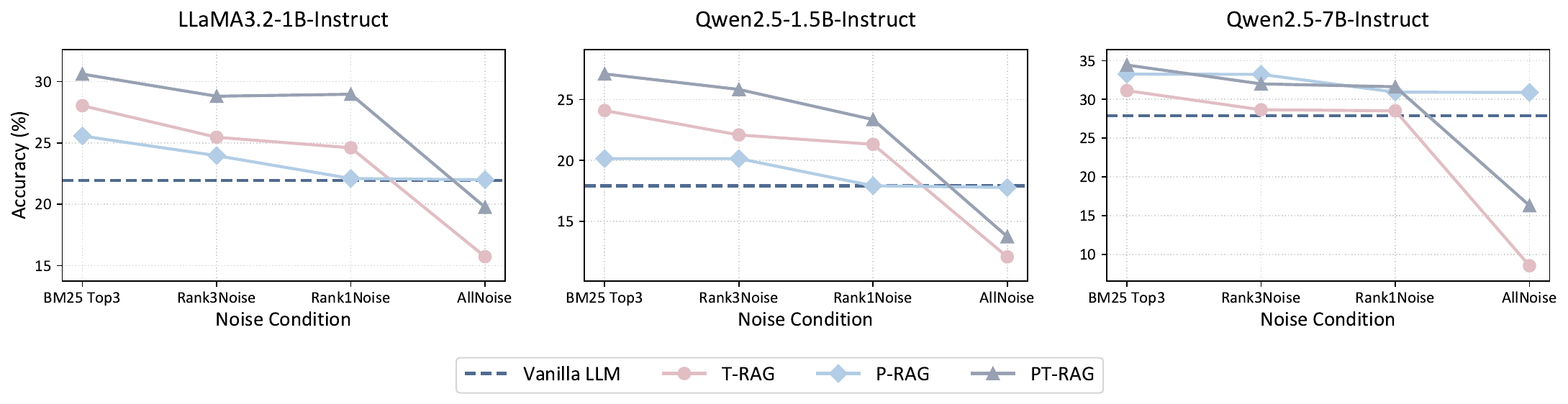}
    \caption{
    Performance trends under increasing levels of retrieval noise.
    The x-axis represents noise severity, ranging from the original retrieved passages (BM25-Top3) to completely irrelevant inputs (AllNoise).}
    \label{fig:noise_results}
\end{figure*}

\heading{Knowledge Conflict}
We evaluate the model's ability to prioritize external evidence over its internal (pre-trained) knowledge in the presence of factual conflicts—a property known as context faithfulness—using the \textbf{ConFiQA} dataset~\cite{bi2024context}.
ConFiQA pairs questions with counterfactual passages and corresponding answers, in which key entities from the original text are replaced with plausible but factually inconsistent substitutes, yielding coherent yet contradictory contexts.
We sample the first 900 instances and use the counterfactual passages for both parametric encoding and inference.
Performance is measured by accuracy against the counterfactual ground-truth answers, directly quantifying the model's capacity to override its internal knowledge in favor of the provided (conflicting) external information.

\heading{Retrieval Noise}
To assess the robustness of parametric RAG against retrieval noise, we introduce controlled artificial noise into the retrieved passages.
Using the same four QA datasets from Section~\ref{sec:repro_setup}, we manipulate the top-3 passages retrieved by BM25 to construct four noise conditions:
\begin{itemize}[leftmargin=*,itemsep=0pt,topsep=2pt,parsep=0pt]
    \item \textbf{BM25-Top3}: The original top-3 BM25-retrieved passages.
    \item \textbf{Rank3Noise}: The 3rd ranked (lowest relevance) passage is replaced with a random irrelevant document.
    \item \textbf{Rank1Noise}: The 1st ranked (highest relevance) passage is replaced with a random irrelevant document.
    \item \textbf{AllNoise}: All three passages are replaced with random noise.
\end{itemize}
We report the average HasAnswer accuracy (as defined in Section~\ref{sec:repro_setup}) across all datasets for each noise condition.

\heading{Cross-Task Generalization}
While \RAG naturally supports diverse tasks via instruction prompting, it remains unclear whether parametric representations—trained exclusively on QA-style objectives—can generalize to other downstream tasks.
We evaluate this transferability on two non-QA tasks using the first 300 instances of each:
(i) \textbf{Fact Verification} using the FEVER dataset~\cite{thorne2018fever}, measured by accuracy; and
(ii) \textbf{Slot Filling} using the Zero-Shot-RE dataset~\cite{levy2017zero}, measured by F1 score.
For each instance, we retrieve the top-3 passages using BM25 and apply the standard \PRAG parameterization pipeline (optimized for answer generation).
However, during inference, we employ task-specific prompts (e.g., ``Judge the claim...'' or ``Predict the slot...'') to assess whether the encoded parameters can adapt to tasks beyond the training objective.

\begin{figure}[tbp]
    \centering
    \includegraphics[width=\linewidth]{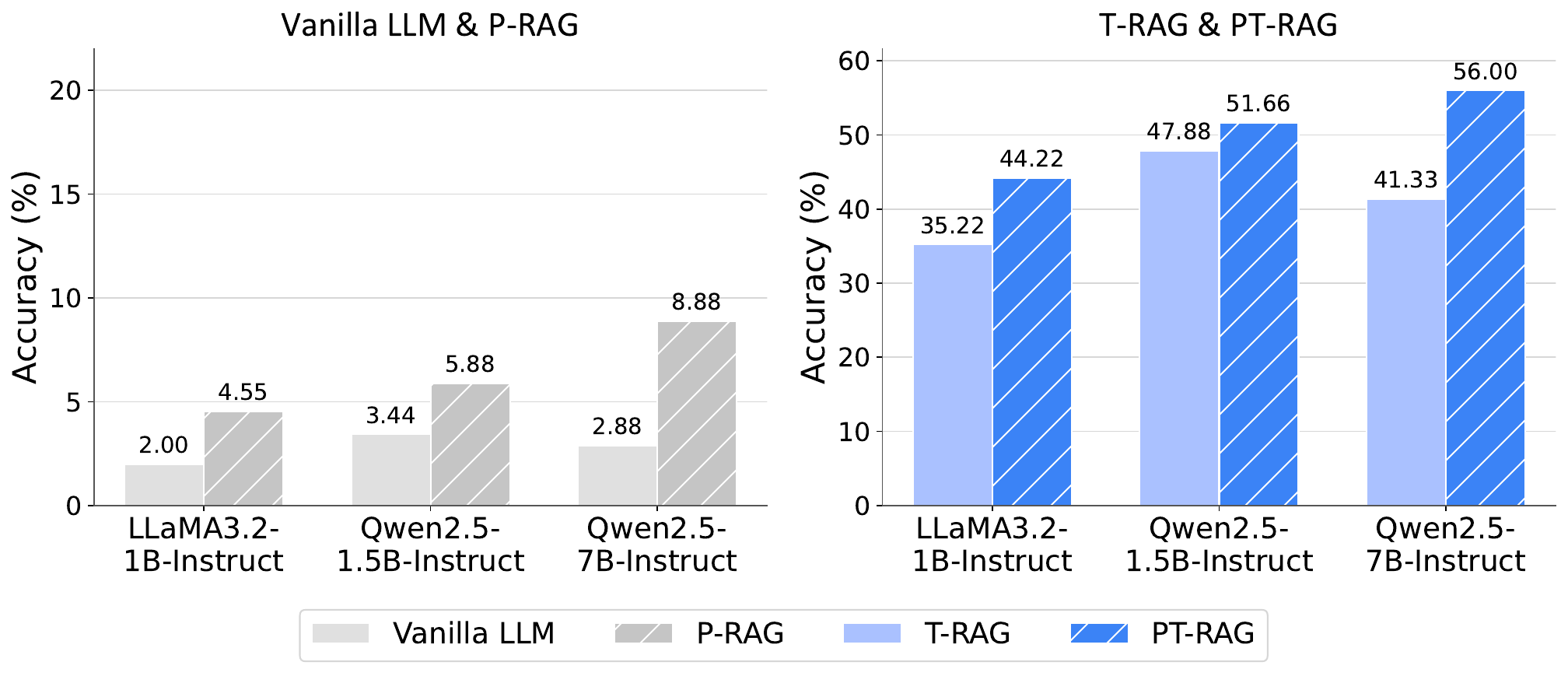}
    \caption{
    Performance on the ConFiQA dataset. Accuracy is computed with respect to the counterfactual ground-truth answers—those consistent with the provided counterfactual context—to evaluate context faithfulness.
    For each model, every pair of methods differs significantly ($p < 0.05$, two-sided paired t-test).}
    \label{fig:cf_results}
\end{figure}

\subsection{Faithfulness under Knowledge Conflicts}

Figure~\ref{fig:cf_results} presents the performance on the ConFiQA dataset.
Our analysis reveals two key observations:
\textbf{(1) \PRAG effectively modifies internal priors.}
Compared to \Vanilla, \PRAG generates significantly more counterfactual answers, confirming that parametric injection can alter the model's pre-trained knowledge.
However, due to the fidelity limitations discussed in Section~\ref{sec:specificity}, 
it still lags behind \RAG in scenarios requiring precise, entity-level knowledge updates.
\textbf{(2) Parametric injection reinforces context faithfulness.}
\PRAGCombine consistently achieves the highest accuracy across all models.
Crucially, this indicates that the injected parameters do not amplify hallucinations; instead, they mitigate the conflict between pre-trained knowledge and the external context, thereby helping the model better adhere to the provided context.

\subsection{Robustness to Retrieval Noise}

Figure~\ref{fig:noise_results} illustrates performance trends under increasing levels of retrieval noise.
Two critical behaviors are observed:
\textbf{(1) Models can discern irrelevant parametric knowledge.}
As the ratio of noisy passages increases, \PRAG's performance gradually declines. 
However, even in the AllNoise setting—where all injected parameters encode entirely irrelevant information—\PRAG never underperforms the \Vanilla baseline.
This indicates that the model is capable of detecting and effectively disregarding unhelpful parametric knowledge, avoiding performance degradation when the injected parameters lack relevance to the query.
\textbf{(2) Parametric injection enhances robustness to context noise.}
\PRAGCombine consistently outperforms \RAG across all noise conditions.
This suggests that parametric injection mitigates the interference from retrieval noise.
By internalizing the document content, the model is better equipped to handle noisy context passages, reducing the distraction caused by irrelevant information during generation.

\subsection{Generalization to Non-QA Task}
\begin{figure}[tbp]
    \centering
    \includegraphics[width=\linewidth]{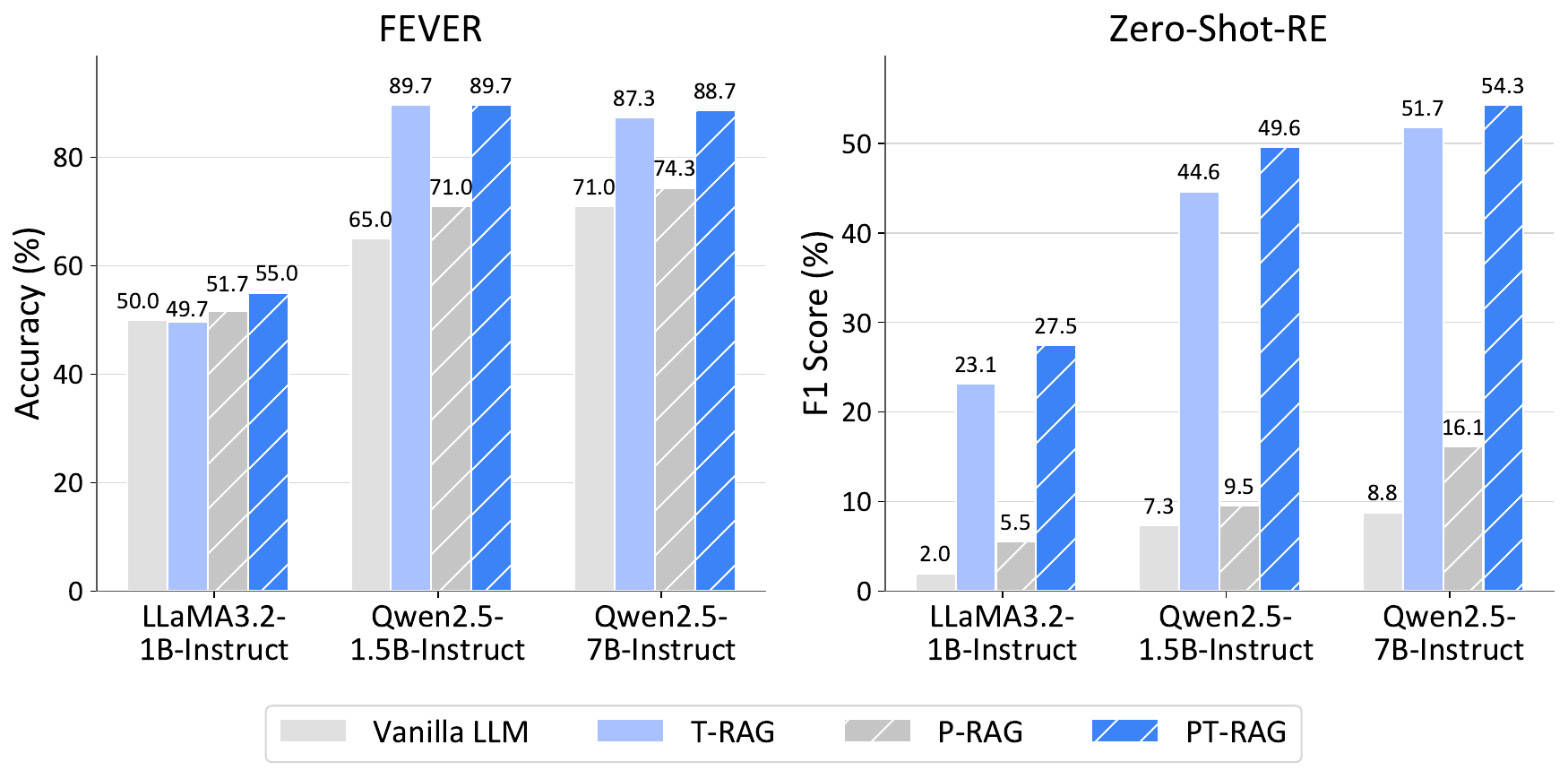}
    \caption{
    Performance on non-QA downstream tasks: Fact Verification (FEVER) and Slot Filling (Zero-Shot-RE).}
    \label{fig:other_task}
\end{figure}

Figure~\ref{fig:other_task} presents the performance on Fact Verification and Slot Filling.
We observe trends consistent with the QA experiments: \PRAG improves over \Vanilla but generally lags behind \RAG, particularly on Slot Filling, a task demanding high-fidelity entity reproduction.
However, \PRAGCombine consistently achieves the best performance.
These results reinforce our earlier conclusion that injected parameters encode high-level document semantics, which are transferable to diverse downstream tasks beyond the specific training objective.

\subsection{Discussion: Practical Challenges}
\label{sec:discussion}

While parametric knowledge injection offers several advantages, our study reveals significant practical barriers that currently hinder its scalability and deployment.
(1) The storage cost for document-specific parameters is prohibitive.
For models employing SwiGLU activations~\cite{shazeer2020glu} (e.g., LLaMA, Qwen), applying LoRA to the FFN layers involves three projections. The parameter count for a single document is calculated as: $P_{\text{doc}} = 3 \cdot L \cdot r \cdot (d_{\text{model}} + d_{\text{ffn}})$,
where $L$ is the number of layers, $r$ is the rank, and $d_{\text{model}}, d_{\text{ffn}}$ denote the hidden and intermediate dimensions.
Taking Qwen2.5-1.5B ($L=28, d_{\text{model}}=1536, d_{\text{ffn}}=8960$) with $r=2$ as an instance, a single parametric representation contains approximately \textbf{1.76M} parameters.
In FP32 precision, this requires $\sim$\textbf{6.73 MB} per document.
Scaling this to just 900 documents consumes nearly \textbf{6 GB}—orders of magnitude larger than the vector indices used in \RAG.
(2) Although \PRAG theoretically reduces inference time complexity by shortening the input context, current serving systems lack optimized support for the high-frequency ``hot-swapping'' of LoRA adapters.
In practice, the system-level latency introduced by loading and managing distinct adapters outweighs the theoretical speedup gained from reduced context length, rendering it slower than standard \RAG in high-throughput scenarios.


\section{Conclusion and Future Work}
This work presents a systematic analysis of parametric RAG.
We find that while parametric injection provides valuable high-level semantic guidance—enhancing multi-hop reasoning, context faithfulness, and robustness to noise—it suffers from significant information loss, falling short of the fidelity offered by explicit token-based context.
These findings underscore that parametric RAG, in its current form, is not a drop-in replacement for conventional RAG, but rather a complementary approach with distinct trade-offs. 
%
Future work could focus on:
(1) optimizing the parameterization pipeline, including training objectives, LoRA configurations, and parameter merging strategies
(2) improving encoding fidelity to mitigate information loss during parameterization;
(3) reducing training and storage overhead through more efficient strategies such as hypernetworks or LoRA decomposition; and
(4) designing systems capable of efficiently loading and managing LoRA adapters at scale.

\clearpage

\bibliographystyle{ACM-Reference-Format}
\balance
\bibliography{reference}


\end{document}